\documentstyle[aps,prl,floats,epsfig]{revtex}
\draft

\begin{document}

\twocolumn[
\hsize\textwidth\columnwidth\hsize\csname @twocolumnfalse\endcsname

\titlepage
\title{Spin alignments of vector mesons in deeply inelastic
lepton-nucleon scattering}
\author {Xu Qing-hua and Liang Zuo-tang }
\address 
{Department of Physics,
Shandong University, Ji'nan, Shandong 250100, China}

\maketitle

\begin{abstract}
We extend the calculations of the spin alignments of vector mesons
in $e^+e^-$ annihilation in a recent Rapid Communication
to deeply inelastic lepton-nucleon scatterings. 
We present the results for different mesons in the current
fragmentation regions of
$\mu^- N$$ \to$$ \mu^- VX$ at high energies
and $\nu_\mu N $$\to$$ \mu^- VX$ 
at both high and low energies.
We also present the predictions for 
$\nu_\mu N$$ \to$$ \mu^- VX$ at NOMAD energies
in the target fragmentation region using a valence quark model.

\end{abstract}

\vspace{0.1in} ]


Spin alignments of vector mesons in high energy reactions 
have attracted much attention recently [1-\ref{XLL01}].
Since the influences from decay of
heavier hadrons are relatively small
and the production rate is in general higher than that of hyperon,
such studies provide important information for the
spin effects in the fragmentation process,
in particular the spin transfer from the fragmenting
quark to the produced hadrons.
Measurements have been carried out in different
reactions [1-\ref{Kress01}] in particular 
in $e^+e^-$ annihilation at LEP recently \cite{Kress01}.
The data show that 
the vector mesons produced in $e^+e^-$ annihilation at $Z^0$ pole
have a large probability to be in the helicity zero state,
and the effect is more significant for large momentum fraction region.
 
In a recent paper\cite{XLL01}, we calculated 
the spin density matrix of vector
mesons in $e^+e^-$ annihilation at $Z^0$ pole by
taking the spin of vector meson 
which contains the fragmenting quark as the 
sum of the spin of the polarized fragmenting 
quark (antiquark) and that of the
antiquark (quark) created in the fragmentation process.
Compared with the data\cite{Kress01}, we showed that
the experimental results for $\rho_{00}$, i.e., 
the probability for vector meson in the helicity zero state, 
imply a significant polarization for the antiquark (quark)
which is created in the fragmentation process and combines with the 
fragmenting quark to form the vector meson.
It should be polarized in the opposite direction
as that of the fragmenting quark and 
the polarization can approximately be written as,
\begin{equation}
P_z = - \alpha P_f ,
\label{eq1}
\end{equation}
where $\alpha \approx 0.5$ is a constant
for most of the vector mesons.
(Here, $P_z$ is the polarization of the antiquark in the moving direction
of the fragmenting quark and $P_f$ is the longitudinal polarization
of the fragmenting quark of flavor $f$) 
Using this result, we were able to fit the data of 
$\rho_{00}$'s for different vector mesons reasonably. 
The relation given by Eq. (\ref{eq1}) can be considered as a 
direct implication of the data\cite{Kress01}
in $e^+e^-$ annihilation.
It should be interesting to extend the
studies to other reactions 
in particular to check whether the relation
shown in Eq. (1) is also true for the fragmentation of 
quarks in other processes.
In this connection, it is encouraging to see that
not only some previous data are available but also
new measurements \cite{DN} can be made by NOMAD collaboration in 
$\nu_\mu N$$ \to$$ \mu^- HX$.
 
In this paper, we extend the calculations
to deeply inelastic lepton-nucleon scatterings (DIS)
and present the results for $\rho_{00}$'s
of different mesons in different cases.
We now start our calculations by summarizing the main points
of the method in Ref. [\ref{XLL01}].
To calculate the density matrix of vector mesons
which are produced in the fragmentation of a polarized quark $q^0_f$,
we divide them into the following two groups 
and consider them separately:
(a) those which contain the fragmenting quark $q^0_f$;
(b) those which don't contain the fragmenting quark.
The spin density matrix $\rho^V(x_F)$ for the vector meson $V$
is given by:

\begin{equation}
\rho^V(x_F)=
\sum_f
\frac {\langle n(x_F|a,f)\rangle}{\langle n(x_F)\rangle}
\rho^V(a,f)
+\frac {\langle n(x_F|b)\rangle}{\langle n(x_F)\rangle}
\rho^V(b)
,
\label{eq2}
\end{equation}
where $\langle n(x_F|a,f)\rangle$ and $\rho^V(a,f)$
are the average number and spin density matrix of vector mesons from (a);
$\langle n(x_F|b)\rangle$ and $\rho^V(b)$ are those from (b);
$\langle n(x_F)\rangle$=
$\sum_f \langle n(x_F|a,f)\rangle$+
$\langle n(x_F|b)\rangle$ is the
total number of vector mesons
and $x_F$ is defined as $x_F$$\equiv$$2p_z^*/W$,
$p_z^*$ is the momentum of vector meson in the $z$ direction 
in the center-of-mass system of total hadronic system, 
$W$ is the total energy of the hadronic system,
and the $z$-axis is taken as the direction of the intermediate boson.
The average numbers $\langle n(x_F|a,f)\rangle$
and $\langle n(x_F|b)\rangle$
are determined by the hadronization mechnism and can be
calculated using hadronization models\cite{lund} as implemented by
the Monte-Carlo event generators.

The vector mesons from group (b) are taken as unpolarized,
thus $\rho^V(b)$=1/3.
For those from group (a) which contain $q^0_f$ and 
an antiquark $\bar q$ created in the fragmentation \cite{note1},  
the spin density matrix $\rho^V(a,f)$ is calculated
from the direct product of the spin density matrix
$\rho^{q^0_f}$ for $q^0_f$ 
and  $\rho^{\bar{q}}$ for $\bar{q}$.
Transforming the direct product, 
$\rho^{q^0_f{\bar q}}$=$\rho^{q^0_f}$$\otimes$$\rho^{\bar q}$,
to the coupled basis 
$|s, s_z\rangle$ (where $\vec{s}$=${\vec {s}^q}$+${\vec {s}^{\bar q}}$),
we obtain the spin density matrix $\rho^V(a,f)$,
and the 00-component as,
\begin{equation}
\rho_{00}^V(a,f)=(1-P_fP_z)/(3+P_fP_z),
\label{eq3}
\end{equation} 
where $P_f$ is the longitudinal polarization of $q^0_f$
and $P_z$ is the polarization of $\bar q$ in $z$-direction. 
We insert the relation showed by Eq. (\ref{eq1}) 
into Eq. (\ref{eq3}) and obtain
\begin{equation}
\rho_{00}^V(a,f)=
{(1+\alpha P_f^2)}/{(3-\alpha P_f^2)}
.               
\label{eq4}
\end{equation}
Finally, from Eqs. (\ref{eq2}) and (\ref{eq4}), we have
\begin{equation}
\rho_{00}^V(x_F)=\sum_f
\frac{1+\alpha P_f^2}{3-\alpha P_f^2}
\frac {\langle n(x_F|a,f)\rangle}{\langle n(x_F)\rangle}
+ \frac{1}{3}
\frac {\langle n(x_F|b)\rangle}{\langle n(x_F)\rangle}
.
\label{eq5} 
\end{equation}
Using this result,
we obtained \cite{XLL01} a good fit to the data of $\rho_{00}^V$'s  
in $e^+e^-$ annihilation.
Now, we apply it to the DIS processes.

We first consider the process of $\mu^-p\to \mu^-VX$.  
For a sufficiently large $W$, 
the hadrons produced in the current fragmentation region are
mainly from the struck quark's fragmentation.
Considering only the leading order subprocess $\mu^-q \to \mu^-q$, 
the polarization of the outgoing struck quark can be obtained
from QED and
can be found in different 
publications [see,e.g.,\ref{LXL01}], i.e.,
\begin{equation}
P_{f}=\frac{P^{l} D_L(y) q_f(x)+P^{N} \Delta q_f(x)}
{q_f(x) +P^{l} D_L(y) P^{N} \Delta q_f(x)},
\label{pqf}
\end{equation}    
where $D_L(y)$ is the longitudinal depolarization factor
$D_L(y)=[1-(1-y)^2]/[1+(1-y)^2]$ and $y\equiv p\cdot (k-k')/p\cdot k$
where $p$, $k$, and $k'$ are the four momenta of the incoming
$q$, $\mu ^-$, and the outgoing $\mu^-$ respectively.
$P^{l}$ and $P^{N}$ are the longitudinal polarizations
of the incoming lepton and nucleon respectively,
$q_f(x)$ and $\Delta q_f(x)$
are the unpolarized
and longitudinally polarized distribution functions.
In our calculation, the unpolarized 
and longitudinal polarized distribution functions
are taken as the GRV98 LO set \cite{GRV98} and the
standard LO scenario of
GRSV2000 \cite{GRSV2000} respectively.

We use the generator {\sc lepto}\cite{lepto} to calculate 
the average numbers
$\langle n(x_F|a,f)\rangle$ and $\langle n(x_F|b)\rangle$.
As an example,
we show the different orgions of $K^{*+}$ 
in the current fragmentation region of 
${\mu^-} N$$ \to$$ \mu^- K^{*+}X$
at the beam energy of 500 GeV in Fig. 1.
We choose the events for
$Q^2$$>$5 (GeV/c)$^2$, $10^{-4}$$<$$x$$<$0.2, 
and 0.5$<$$y$$<$0.9 to ensure a 
reasonably large $W$ and a reasonably high polarization of
the fragmenting quark.
We see that, in contrast to hyperon production in
the same reaction \cite{LXL01}, the decay contribution
is indeed very small.

Using Eq.(\ref{eq5}) and the average numbers obtained above,
we calculate the $\rho^V_{00}$'s
for $K^{*0}$, $K^{*+}$, $\rho^{\pm}$ and $\rho^{0}$. 
The results are shown in Fig. 2
for different combinations of $P^{l}$ and $P^{N}$
in the same kinematic region as in Fig. 1.
We see that $\rho^V_{00}$'s increase with increasing $x_F$ and 
reach about 0.5 in the cases of polarized lepton beam,
which is much larger than $1/3$
and can easily be detected in experiments. 
However, the magnitude of $\rho^V_{00}$'s for $P^{l}$=0 and $P^{N}$=1 
is much smaller than those in other three cases.
The reason is that the ratio $\Delta q_f(x)/q_f(x)$ is small
in the chosen small $x$ region, so $P_f$ 
obtained from Eq.(\ref{pqf}) is small. 
While the large $D_L(y)$ for large $y$ region
leads to relative large $P_f$ in other three cases, 
the result of which is large $\rho^V_{00}$.
Hence, to get reasonably large $\rho^V_{00}$,
the polarization of the lepton beam is required.   

For inclusive meson production in unpolarized $ep$ reactions, 
the struck quark is unpolarized.
Hence, $P_f$=0 and $\rho_{00}^V$=1/3
in the current fragmentation region.
There have been measurements \cite{epr82} for 
$\rho^{0}$ with $z$$>$0.4
($z$=$E_{\rho}/\nu$ and $\nu$ is the energy loss
of the lepton beam in the lab frame) 
in inelastic $ep$ scattering at $E_e$=11.5 GeV
and the result is $\rho_{00}$=0.41$\pm$0.08,
which is in agreement with the theoretical expectation.

For $\nu_{\mu} N$$ \to$$ \mu^- VX$, the leading
subprocess $\nu_\mu q$$ \to$$ {\mu}^- q$ is a charged current weak 
interaction with the exchange of a virtual $W^+$,
which selects only left-handed quarks or right-handed antiquarks.
Thus, $P_f=-1$ for struck quarks
and $P_f$=1 for struck antiquarks,
whose polarizations reach the maxmum.
Hence,
we expect larger $\rho^V_{00}$'s  in neutrino DIS
than in other reactions.
We calculate $\rho^V_{00}$'s for different mesons 
in the current fragmentation region of $\nu_{\mu} N$$\to$$ \mu^- VX$
at $E_{\nu}$=500 GeV and
the results are shown in Fig. 3. 
We see that, the $\rho^V_{00}$'s for $K^{*+}$, 
$\rho^{+}$ and $\rho^{0}$ increase to
about 0.6 with increasing $x_F$.
However,
the $\rho_{00}$ for $K^{*0}$ is much smaller.
This is because, for $K^{*0}$,
the only contribution of type (a) is from outgoing 
struck $\bar s$ quark which is a result of the absorption
of $W^+$ by a $\bar c$ in the nucleon sea.
It is very small due to the
rarity of $\bar c$ in the nucleon.
Other kinds of contributions lead only to 1/3 for $\rho_{00}$.
For other vector mesons, such as $K^{*-}$ and $\rho^{-}$, 
there is no contribution of type (a) at all.
Their $\rho_{00}$'s are just equal to 1/3.

It has been pointed out that \cite{LL02},
at lower energies, such as 
at the NOMAD energies where $\langle E_\nu \rangle$=43.8 GeV,
the influence of the fragmentation of target remnant
to hyperon production is very large,
in particular in the small $x_F$ region \cite{LL02}.
It has to be taken into account in calculating the
hyperon polarization in such energy region.
The characteristic features of the hyperon polarization
in this region are determined by this contribution.
It is therefore natural to ask whether similar effects
also exist for mesons production.
To check this,
we make an analysis using {\sc lepto}
and the results for 
$\nu_{\mu} p$$ \to$$ \mu^- K^{*+}X$ at $E_\nu$=43.8 GeV
are shown in Fig. 4.
We see that, there is indeed a 
mixture of the contribution from the mesons containing the
struck quark and that containing one of the quarks in 
the target remnant in the region near $x_F$$\approx$0,
but the effect is much smaller than that for hyperon production.
This can be understood easily.
We recall that, in the case of hyperon production,
the excitation of diquark-anti-diquark pair is needed
for producing hyperons which contain the struck quark
and there should be at least one more baryon and 
one antibaryon produced.
However, the excitation of diquark-anti-diquark pair is 
unnecessary for producing hyperons which contain one quark of the 
remnant $uu$ diquark. 
At the NOMAD energies, $W$ is only of several GeV,
the probability for the former case
should be much smaller than that for the latter.
Hence, in the $x_F\sim 0$ region,
the contributions from the latter case can dominate.
For meson production,
to produce mesons containing the struck quark or
one $u$ quark of the remnant $uu$ diquark,
only quark-antiquark pair excitation is needed.
The probability for the mesons containing one quark of $uu$ diquark
to move in the opposite direction
of the diquark is rather small.
Hence, the influences from target remnant
fragmentation on the spin alignments of 
vector mesons in current fragmentation region are
small.
  
Having the above-mentioned results for meson production of
different origins,
we can also calculate their spin alignments
by taking the fragmentation of the nucleon remnant into account.
The polarization of the quarks in target remnant is unclear.
We calculate it in the same way as in Ref.[\ref{LL02}] 
in studying hyperon polarization,
where a valence quark model was used.
The results obtained for $\rho^V_{00}$'s
in $\nu_{\mu} p$$ \to$$ \mu^- VX$ at $E_\nu$=43.8 GeV
both in the current and target fragmentation regions
are shown in Fig. 5.
We see that, for $x_F>0$,
the influence from the target remnant fragmentation
is indeed very small.
Compared with those obtained at $E_\nu$=500 GeV,
$\rho_{00}$ of $K^{*0}$ is smaller.
This is because at such low energy, the probability for
the outgoing struck quark to be $\bar s$ is very tiny.
For the target fragmentation region,
the spin alignments are smaller than those in the current region.
The results for a neutron target are shown in Fig. 6. 
The $\rho^V_{00}$'s are much smaller in the 
target fragmentation region
than those in case of a proton target,
because the polarization of the quark in the remnant $uu$ diquark
from a proton is larger than that 
in the remnant $ud$ diquark from a neutron \cite{LL02}.
The difference in the current region is 
tiny for the two different targets,
since the partonic subprocesses are the same 
and $\nu_{\mu} d$$ \to$$ \mu^- u$ dominates others
for both targets. 

There have been measurements \cite{BEBC87}
for $\rho^0$ in neutrino DIS and the results 
are $\rho_{00}$=0.65$\pm$0.18 and 0.61$\pm$0.08
in $\bar{\nu}N_e$$ \to$$ \mu^+ \rho ^0X$
and $\rho_{00}$=0.41$\pm$0.13 and 0.39$\pm$0.08
in ${\nu}N_e$$ \to$$ \mu^- \rho ^0X$ at low energies similar to NOMAD.
The data are both for $\rho^0$'s with $z$$>$0.4 ($z$=$E_{\rho}/\nu$).
Our results in the same kinematic region
are $\rho_{00}$=0.511 for $\bar{\nu}N_e$$ \to$$ \mu^+ \rho ^0X$
and $\rho_{00}$=0.518 for ${\nu}N_e$$ \to$$ \mu^- \rho ^0X$ \cite{note}.
They are in agreement with the data
and further measurements with high precision are required
to give a better check of the model.

In summary,
we calculate the spin alignments for different
vector mesons
in the current fragmentation regions of 
$\mu^- N$$ \to$$ \mu^- VX$ and $\nu_\mu N$$ \to$$ \mu^- VX$
at high energies by taking the spin of a vector meson
as the sum of the spins of the polarized fragmenting
quark (antiquark) and that of the
antiquark (quark) created in the fragmentation process.  
We also present the predictions for spin alignments
in $\nu_\mu N$$ \to$$ \mu^- VX$ at NOMAD energies
both in the current and target fragmentation regions.
The results show that there are significant spin alignments
for most of the vector mesons in the above reactions.
Measurements of them can provide important information for the
spin effects in the fragmentation process.

We thank Li Shi-yuan, Liu Chun-xiu, Xie Qu-bing
and other members in the theoretical particle physics group of
Shandong University for helpful discussions.
This work was supported in part by the National Science Foundation
of China (NSFC) and the Education Ministry of China
under Huo Ying-dong Foundation.

\noindent



\begin{figure}[h]
\psfig{file=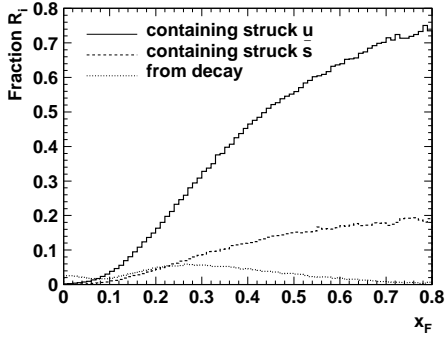,width=7cm}
\caption{Different contributions to $K^{*+}$ production
in the current region of 
${\mu^-} N$$ \to$$ \mu^- K^{*+}X$
at $E_\mu$=500 GeV. 
}
\label{fig2}
\end{figure}
 
\begin{figure}[h]
\psfig{file=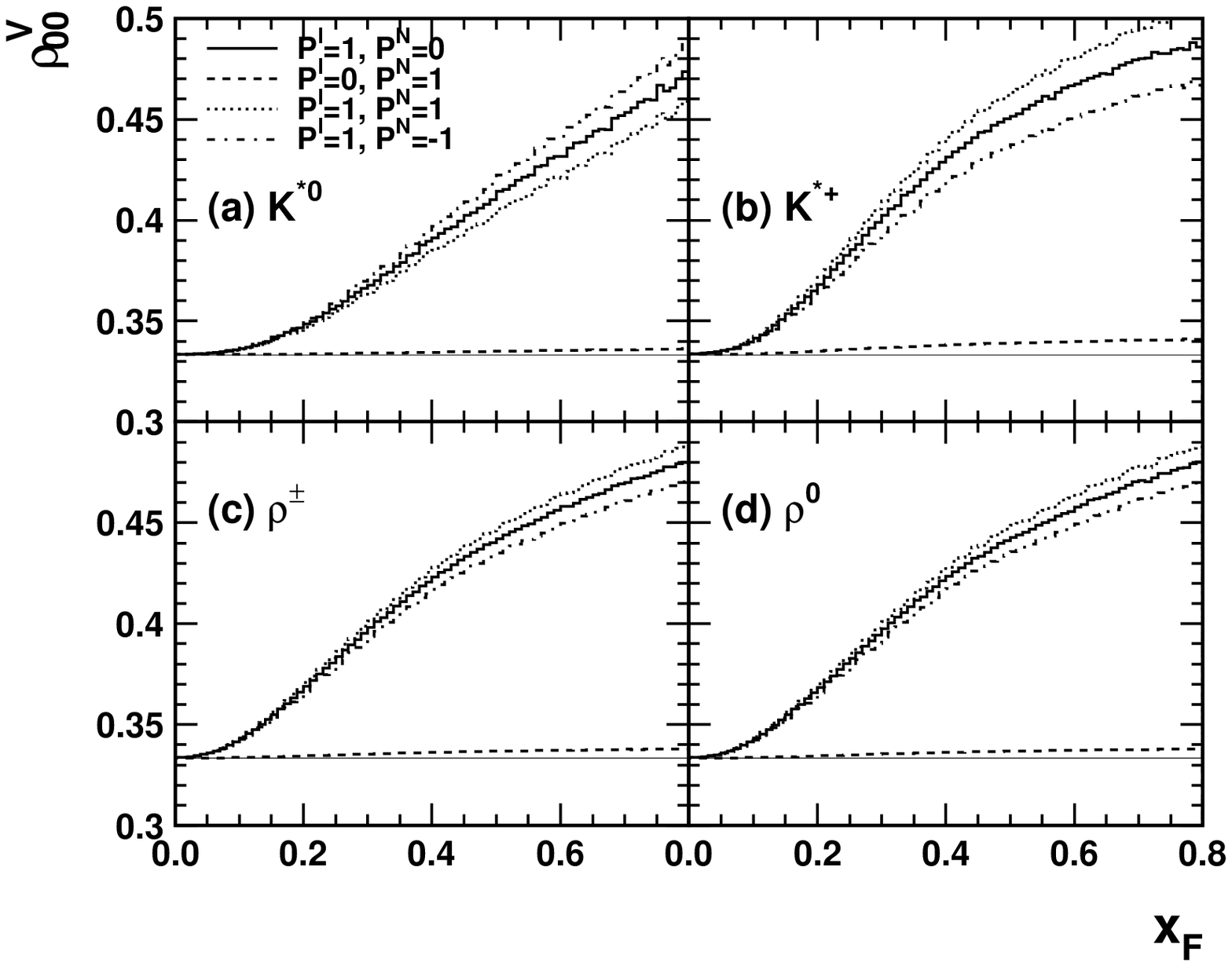,width=8cm}
\caption{$\rho^V_{00}$ in the current
 region of $\mu^- p$$ \to$$ \mu^- VX$ at $E_\mu$=500 GeV.  
The straight lines at $\rho_{00}$=1/3 show
the unpolarized cases.}
\label{fig3}
\end{figure}

\begin{figure}[b]
\psfig{file=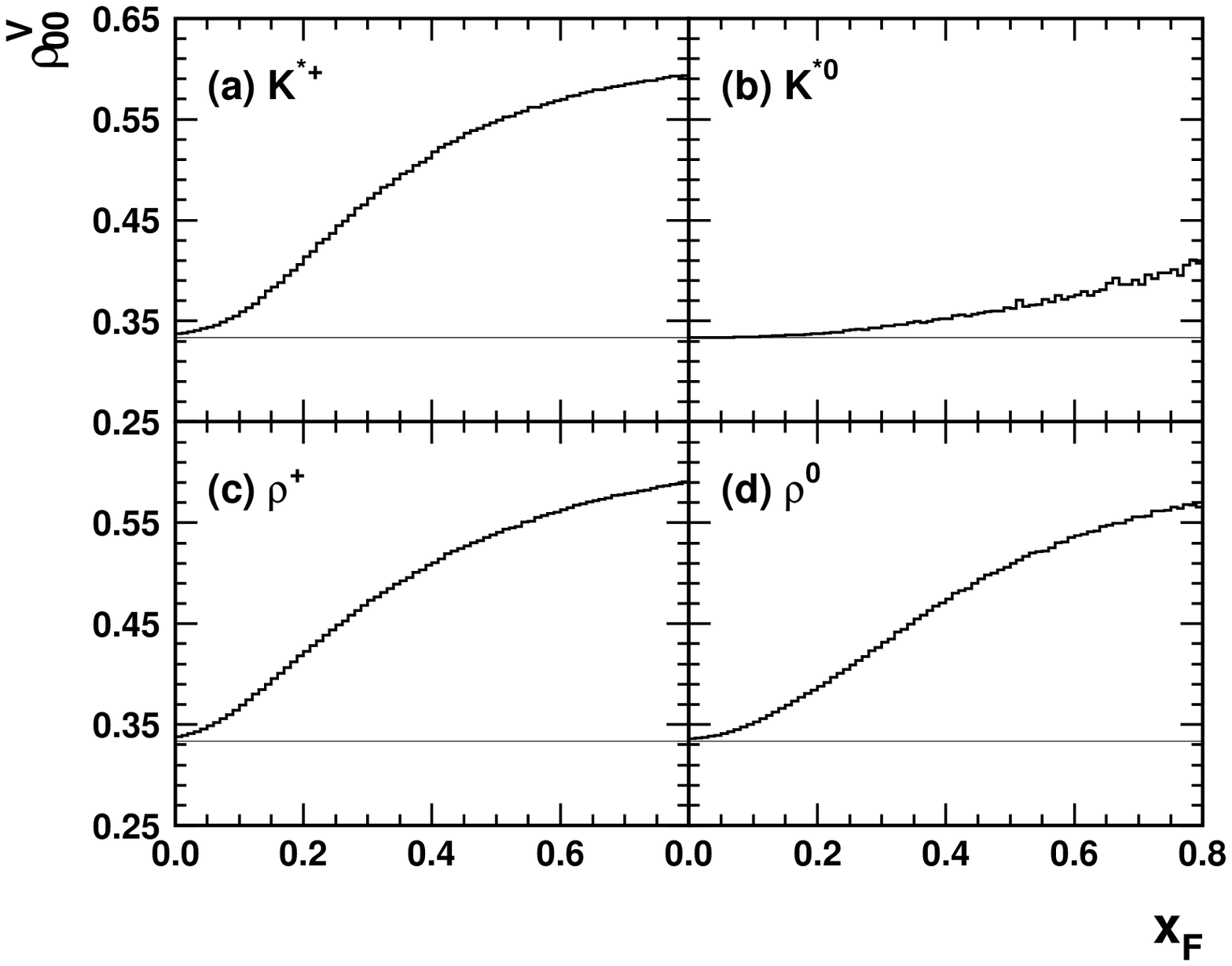,width=8cm}
\caption{$\rho^V_{00}$ in the current
region of $\nu_\mu p$$ \to$$ \mu^- VX$ at $E_\mu$=500 GeV.
The straight lines at $\rho_{00}$=1/3 show
the unpolarized cases.}
\label{fig4}
\end{figure}

\begin{figure}[h]
\psfig{file=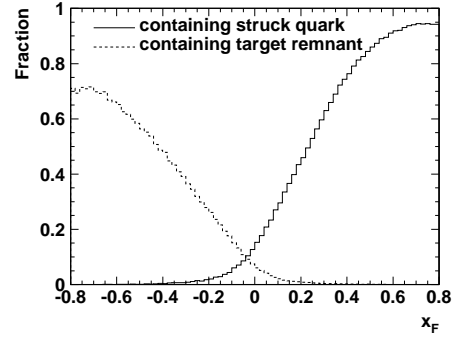,width=7cm}
\caption{Different contributions to $K^{*+}$ production both in the
current and target fragmentation region of
$\nu_\mu p$$ \to$$ \mu^- K^{*+}X$ at $E_\nu$=43.8 GeV.
}
\label{fig5}
\end{figure}

\begin{figure}[h]
\psfig{file=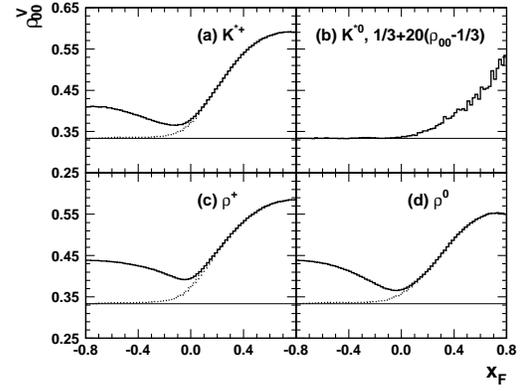,width=8cm}
\caption{$\rho^V_{00}$ in
$\nu_\mu p$$ \to$$ \mu^- VX$ at $E_\nu$=43.8 GeV. 
The solid line represents the results where the contribution of
target fragmentation is taken into account,
the dotted line denotes the results where only the contribution
of the current fragmentation is included.
The results for $K^{*0}$ are same.
 }
\label{fig6}
\end{figure}

\begin{figure}[h]
\psfig{file=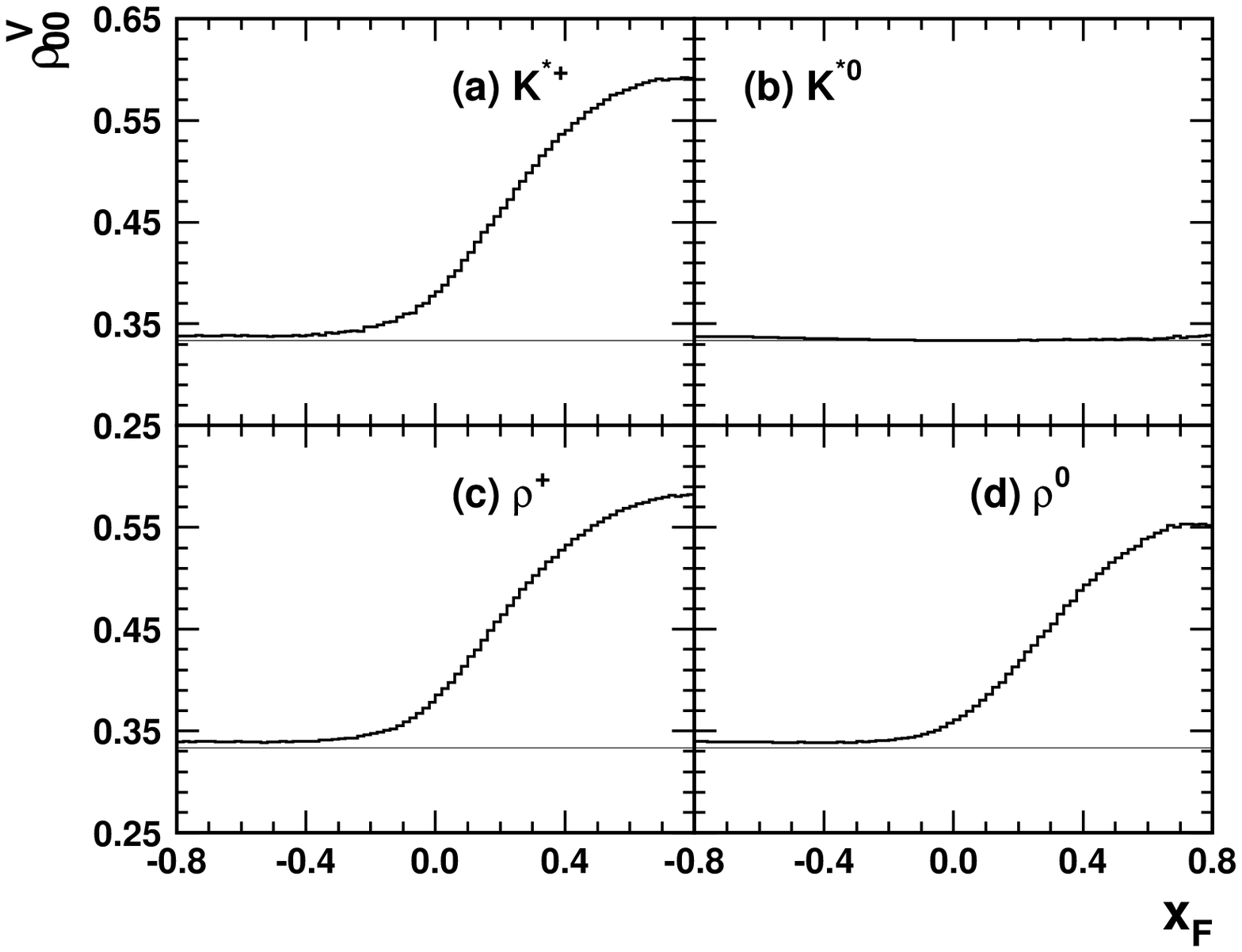,width=8cm}
\caption{$\rho^V_{00}$ in
$\nu_\mu n$$\to$$\mu^- VX$ at $E_\nu$=43.8 GeV.  }
\label{fig7}
\end{figure}


\begin {thebibliography}{99}
\bibitem{epr82} I. Cohen. et al., Phys. Rev. D 25, 634, (1982).
\bibitem{BEBC87} BEBC WA59 Collab., W. Witttek et al.,
 Phys. Lett. {\bf B187}, 179 (1987);
 BBCN Collab., V.G. Zaetz et al.,
  Z. Phys. {\bf C66}, 583 (1995).
\bibitem{EXCHARM00} EXCHARM Collab., A. N. Aleev et al.,
 Phys. Lett. {\bf B485}, 334 (2000).
\bibitem{Kress01}
 DELPHI Collab., P. Abreu et al.,
 Phys. Lett. {\bf B406}, 271 (1997);
 OPAL Collab., K. Ackerstaff et al.,
 Phys. Lett. {\bf B412}, 210 (1997);
 OPAL Collab., K. Ackerstaff et al.,
 Z. Phys. {\bf C74}, 437 (1997);
 OPAL Collab., G. Abbiendi et al.,
 Eur. Phys. J. {\bf C16}, 61 (2000). 
\label{Kress01}
\bibitem{Falk94} A.F. Falk and M.E. Peskin, 
 Phys. Rev. D {\bf 49}, 3320(1994).
\bibitem{Ansel} M. Anselmino, M. Bertini, F. Murgia and B. Pire, 
  Phys. Lett. {\bf B438}, 347 (1998);
  M. Anselmino, M. Bertini, F. Murgia and P. Quintairos, 
 Eur. Phys. J. {\bf C11}, 529 (1999).  
\bibitem{JPMa02} J. P. Ma, Nucl. Phys. B{\bf 622}, 416 (2002). 
\bibitem{XLL01} Xu Qing-hua, Liu Chun-xiu and Liang Zuo-tang,
 Phys. Rev. D 63, 111301(R) (2001).
\label{XLL01}
\bibitem{DN} D. Naumov, private communication (2001).
\bibitem{lund} B.~Andersson, 
 G.~Gustafson, G.~Ingelman, and T.~Sj\"ostrand,  
 Phys. Rep. {\bf 97}, 31 (1983).
\bibitem{note1}
It is similar for those of type (a) from the fragmentation of
a polarized antiquark $\bar{q}^0_f$,
i.e., $V$=$({\bar q}^0_f q)$.
\bibitem{LXL01} Liu Chun-xiu, Xu Qing-Hua, and Liang Zuo-tang,
 Phys. Rev. {\bf D 64}, 073004 (2001).
\label{LXL01}
\bibitem{GRV98} M. Gl\"uck, E. Reya, and A. Vogt,
 Eur. Phys. J. {\bf C 5}, 461 (1998). 
\bibitem{GRSV2000} M. Gl\"uck, 
 E. Reya, M. Stratmann, and W. Vogelsang,
 Phys. Rev. {\bf D 63}, 094005 (2001).
\label{GRSV2000}
\bibitem{lepto} G. Ingelman, A.Edin, J.Rathsman, 
        Comp. Phys. Comm. {\bf 101}, 108 (1997).
\bibitem{LL02} Liang Zuo-tang and Liu Chun-xiu, hep-ph/0204323,
 submitted to Phys. Rev. {\bf D}; See also, Liang Zuo-tang,
 talks at 31st Int. Sym. on Mitiparticle-dynamics (ISMD31)
 and 3rd Circum-Pan-Pacific Symposium on High Energy Spin Physics,
 hep-ph/0111403 and hep-ph/0205017.
\label{LL02}
\bibitem{note}
We note that the quantization axis of the data is chosen
along the moving direction of the vector meson
(helicity frame),
while in our calculations it is chosen as the moving
direction of the outgoing struck quark.
There is a slight difference between the results in
these two frames \cite{XLL01}
and the magnitudes of the differences
depend on the transverse momentum of the meson with respect to
the outgoing struck quark and other related factors.
We estimate that,
typically, the correction is within 0.05. 


\end{thebibliography}
\end{document}